\documentstyle[twocolumn]{mn}
\input psfig
\def\lsim{~\rlap{$<$}{\lower 1.0ex\hbox{$\sim$}}}
\def\bsim{~\rlap{$>$}{\lower 1.0ex\hbox{$\sim$}}}

\def\dd{{\rm d}}
\def\ln{{\rm ln}}

\def\pmb#1{\setbox0=\hbox{#1}%
\kern-.025em\copy0\kern-\wd0
\kern.05em\copy0\kern-\wd0
\kern-.025em\raise.0433em\box0}

\def\vv{\pmb{$v$}}
\def\valpha{\pmb{$\alpha$}}
\def\vtheta{\pmb{$\beta$}}

\def\vx{\pmb{$x$}}

\def\vg{\pmb{$g$}}

\def\etal{{\it et al.\ }}

\def\vnabla{\pmb{$\nabla$}}

\title[The $\Omega$ Dependence in the Equations of Motion]
{The $\Omega$ Dependence in the Equations of Motion}

\author[Nusser \& Colberg]{Adi Nusser \& J\"org M. Colberg\\Max Planck
 Institut fur Astrophysiks, Karl Schwarzschild
Str. 2,  D--85748
Garching b. M\"unchen, Germany}

\begin{document}

\maketitle


\begin{abstract}
We show that the equations of motion governing the evolution of a
collisionless gravitating system of particles in an expanding universe
can be cast in a form which is {\it almost} independent of the
cosmological density parameter, $\Omega$, and the cosmological
constant, $\Lambda$. The new equations are expressed in terms of a
time variable $\tau\equiv \ln D$, where $D$ is the linear rate of
growth of density fluctuations.  The weak dependence on the density
parameter is proportional to $\epsilon=\Omega^{-0.2}-1$ times the
difference between the peculiar velocity (with respect to $\tau$) of
particles and the gravity field (minus the gradient of the potential),
or, before shell-crossing, times the sum of the density contrast and
the velocity divergence.  In a 1-dimensional collapse or expansion,
the equations are fully independent of $\Omega$ and $\Lambda$ before
shell-crossing.  In the general case, the effect of this weak $\Omega$
dependence is to enhance the rate of evolution of density
perturbations in dense regions.  In a flat universe with $\Lambda\ne
0$, this enhancement is less pronounced than in an open universe with
$\Lambda=0$ and the same $\Omega$.  Using the spherical collapse
model, we find that the increase of the $rms$ density fluctuations in
a low $\Omega$ universe relative to that in a flat universe with the
same linear normalization is $\sim 0.01 \epsilon(\Omega)
<\!\delta^3\!>$, where $\delta$ is the density field in the flat
universe.  The equations predict that the smooth average velocity
field scales like $\Omega^{0.6}$ while the local velocity dispersion
($rms$ value) scales, approximately, like
$\Omega^{0.5}$. High resolution N-body simulations confirm these
results and show that density fields, when smoothed on scales slightly
larger than clusters, are insensitive to the cosmological model. Halos
in an open model simulation are more concentrated than halos of the
same $M/\Omega$ in a flat model simulation.
\end{abstract}

\begin{keywords}
cosmology: theory -- dark matter -- large scale structure of Universe
\end{keywords}
\section {Introduction}

The cosmological background determines the growth rate of matter
density fluctuations. This is the result of two effects. First, the
initial conditions are specified in terms of the density contrast
field $\delta\equiv\rho(\vx)/\rho_b-1$. Therefore the actual density,
$\rho(\vx)$, which dictates the dynamical evolution, as can be seen
for example from the spherical top-hat model, depends on the mean
matter density, $\rho_b$.  The second effect comes about simply
because the mean matter density varies with time according to the
assumed cosmological model. This in turn translates into a dependence
of the evolution of the fluctuation field $\delta$ on the parameters
of the cosmological model: the density parameter, $\Omega$, and the
cosmological constant, $\Lambda$.  Here we focus on the following
aspect of the dependence of dynamics on the cosmological background.
Starting from an initial density fluctuation field and a given
amplitude of the evolved field, we address the question: how do the
evolved peculiar velocity and density fields depend on the parameters
$\Omega$ and $\Lambda$?  In the linear (e.g. Peebles 1980) and in the
Zel'dovich quasilinear (Zel'dovich 1970) approximations, once an
initial density fluctuation field is evolved to a given amplitude, it
does not contain any information
 on the parameters $\Omega$ and
$\Lambda$. In these approximations, the peculiar velocity field is
simply proportional to $f(\Omega,\Lambda)$ where $f$ is the so-called
linear growth factor.  This result is easy to understand.  In the
linear approximation, the density fluctuations are merely amplified by
a time dependent factor, $D$.  In the Zel'dovich approximation, the
displacement vector is the product of the initial gravity field and
the function $D$.  Moreover, second order perturbation theory
calculations (e.g. Bouchet \etal 1992) have shown that moments of the
density fluctuation field are very insensitive to $\Omega$ and
$\Lambda$.  Finally, in the highly nonlinear regime, N-body
simulations (e.g. Davis \etal 1985) show that the final matter
distribution in simulations with the same initial conditions changes
very little as the parameters of the cosmological background are
varied.  Significant differences between flat and open models are
found only in the cores of what are identified as rich clusters in
these simulations.  These results have proved useful in analyzing
observations of the large scale structure. Nusser and Dekel (1993),
for example, used N-body simulations to argue that a recovery of the
initial density fluctuations from the observed galaxy distribution is
almost independent of $f(\Omega,\Lambda)$, where as a recovery from
the observed peculiar velocity field is sensitive to the assumed
$\Omega$. They applied their reconstruction method to the POTENT
compilation of the peculiar velocity data and to the 1.2 Jy IRAS
survey and concluded that $\Omega>0.3$ with high
confidence. Bernardeau \etal (1995) used second order perturbation
theory to argue that the reduced skewness of the divergence of the
peculiar velocity field is inversely proportional to $f$, in
accordance with the scaling implied by the Zel'dovich
approximation. They found that the measured skewness is consistent
with $\Omega$ of about unity.

Here we aim at a better understanding of the dependence of the
equations of motion on the cosmological parameters. In section 2 we
write the equations of motion in a form which is almost independent of
the background cosmology. We discuss the $\Omega$ dependence in
toy-models in section 3.  In section 4, we use high resolution N-body
simulations to investigate in detail the differences in the matter
distribution in flat and open models. We conclude with a summary and
discussion in section 5.

\section {Almost $\Omega$ independent equations of motion}
We restrict our treatment to the case of a matter dominated
universe with a cosmological constant, i.e, we assume that 
the total mean density
is $\rho_{tot}=\rho_b+\Lambda/3$ where $\rho_b(t)$ is the
matter contribution and $\Lambda $ is the cosmological constant.
We use the standard notation in which,
$a(t)$ is the scale factor, $H(t)=\dot{a}/a$ is the time dependent 
Hubble factor, $\Omega=\rho_b(t)/\rho_c(t)$ and 
$\lambda=\Lambda/3H^2$
where $\rho_c=3H^2/8\pi G$ is the critical density.
Let $\vx$ and $\vv=\dd \vx /\dd t$ be the position and
peculiar velocity of a particle in comoving
coordinates. 
The equations of motion are:
the continuity equation
\begin{equation}
\frac{\dd \delta}{\dd t} + \left(1+\delta\right)\vnabla \cdot \vv
=0 , \label{cont}
\end{equation}
the Euler equation of motion,
\begin{equation}
\frac{\dd \vv}{\dd t} +2H\vv=-\frac{3}{2}\Omega \vnabla \phi , \label{euler}
\end{equation}
and the Poisson equation,
\begin{equation}
\Delta \phi =  \delta . \label{poisson}
\end{equation}
Note that we have defined $\phi\equiv 2\Phi_g/3\Omega$ where
$\Phi_g$ is the peculiar gravity potential in comoving coordinates.
Equations (\ref{cont}), (\ref{euler}) and (\ref{poisson}) together
with the Friedman equations for the background quantities $\Omega$ and
$a$ fully specify the dynamics of pressure free density fluctuations.
The scale factor, $a$, can be solved for using the Friedman equation
\begin{equation}
\left(\frac{\dd a}{\dd t}\right)^2=
\frac{8\pi}{3}G\rho_ba^2+\frac{\Lambda}{3}a^2 -k , \label{scale}
\end{equation}
where $k=+1$, $-1$ and 0  correspond, respectively,
to closed, open or flat universes.
Energy conservation, $\rho_b
a^3=const$,  yields $\Omega=c_0/H^2a^3$
where $c_0=\Omega_0 H_0^2 a_0^3$ and the subscript 0 denotes quantities at the present time.
Therefore, energy conservation and   (\ref{scale}) yield
\begin{equation}
\Omega=\frac{c_0}{c_0
+\frac{\Lambda}{3}a^3+k a} . \label{omev}
\end{equation}

The hubble factor, $H$, can be eliminated from these equations by
working with a new time variable $p\equiv \ln a$.  We define a new
``velocity'' $\valpha\equiv \dd \vx /\dd p=H^{-1}\vv$.  In these new
variables, the Poisson equation remains unaltered while the continuity
and Euler equations become
\begin{equation}
\frac{\dd \delta}{\dd p} + \left(1+\delta\right)\vnabla \cdot \valpha
=0 , \label{contalpha}
\end{equation}
\begin{equation}
\frac{\dd \valpha}{\dd p} +\left(1-q\right)
\valpha=-\frac{3}{2}\Omega \vnabla\phi , \label{euleralpha}
\end{equation}
where  $q(p)=\Omega/2- \lambda$ is
the time dependent deceleration parameter and we have 
used $\dd H / \dd p=-(1+q)H$ to derive
(\ref{euleralpha}).

Attempting to eliminate $\Omega$ from (\ref{euleralpha}) 
we further make an additional transformation from the
time variable $p$ to $\tau$ defined by
\begin{equation}
\tau=\ln D(p) , \label{tau}
\end{equation}
where $D$ is the linear growing density mode determined by the equation,
\begin{equation}
\frac{\dd^2 D}{\dd t^2}+2H\frac{\dd D}{\dd t}-\frac{3}{2}\Omega H^2
D=0 . 
\label{growth}
\end{equation}
Analytic solutions to (\ref{growth}) can be found in Heath (1977).
Expressing (\ref{euleralpha}) in terms of $\tau$ defining the
 ``velocity'' $\vtheta\equiv\dd \vx /\dd \tau$, 
continuity equation is
\begin{equation}
\frac{\dd \delta}{\dd \tau}-(1+\delta)\theta=0 , 
\label{contnom}
\end{equation}
and the Euler equation is
\begin{equation}
f^2{{\dd \vtheta}\over {\dd \tau}}+
\left[{{\dd f}\over {\dd p}}+\left(1-q
\right)f\right]\vtheta
=-{3\over 2}\Omega \vnabla \Phi  , \label{eulertheta}
\end{equation}
where, $\theta=-\vnabla \cdot \vtheta$, and  $f=\dd \tau /\dd p$ is the linear growth factor which relates
density contrast, $\delta$,  to the divergence of the peculiar 
velocity field, $\vv$, in the linear regime.  For $\lambda=0$, a good 
approximation
\footnote{For $\lambda=0$, the function $f$ satisfies
\((1-\Omega)\frac{\dd f}{\dd {\rm ln}\Omega}-(1-\frac{\Omega}{2})f
+\frac{3}{2}\Omega-f^2=0 , \) which to first order in $1-\Omega$
yields $f\approx \Omega^{4/7}$ for $\Omega \approx 1$ (see also
Lightman \& Schechter 1990).  However, the general solution to this
equation is better fitted by $f\approx \Omega^{0.6}$ for $\Omega<0.7$.
A fit which works well for $0.05<\Omega<1$ is \(f=\Omega^{\frac{4}{7}+
\frac{(1-\Omega)^3}{20}} . \)} for $f$ is $f\approx \Omega^{0.6}$
(Peebles 1980).  For $\lambda\neq 0$, Lahav \etal (1991) found that
$f\approx \Omega^{0.6}+\lambda(1+\Omega/2)/70$. Therefore, for
reasonable values of the cosmological constant we neglect the
dependence of $f$ on $\lambda$ in the approximate forms for $f$ (Lahav
\etal 1991).  Note that the velocity $\vtheta=\vv/(Hf)$. In the
Zel'dovich approximation, this is the displacement vector of a particle
from its initial to present position.  Using (\ref{growth}), we find
\begin{equation}
{{\dd f}\over {\dd p}}+\left(1-q\right)f=
\frac{3}{2}\Omega -f^2 , \label{growthf}
\end{equation}
so the Euler equation (\ref{eulertheta}) is 
\begin{equation}
\frac{\dd \vtheta} {\dd \tau}
-\vtheta -\frac{3}{2}\left[1+\epsilon 
(\Omega)\right]\left(\vg -\vtheta\right)=0 , \label{eulernom}
\end{equation}
where
\begin{equation}
\epsilon(\Omega)\equiv\frac{\Omega}{f^2}-1\approx \Omega^{-0.2}-1 ,
\label{epsdef}
\end{equation}
and we have defined $\vg=-\vnabla \phi$.
 The weak dependence on $\Omega$ in 
(\ref{eulernom}) through $\epsilon(\Omega)$ couples
to the difference between the velocity, $\vtheta$, and the gravity
field, $\vg$.
Since $\Omega \sim 1$ at early times, initially the
function  $\epsilon$ almost
vanishes, thus, any changes in the dynamics as a result of
this weak dependence on $\Omega$ occur at later times.

In virialised regions, the acceleration of a particle is dominated by
the gravity field $\vg$.  It is easy to see that by neglecting the
terms involving the velocity in (\ref{eulernom}) and working with a
new time variable with respect to which the velocity is
$(1+\epsilon)^{-1/2}\vtheta\approx \Omega^{0.1}\vtheta $ we obtain an
Euler equation which is independent of $\Omega$. This velocity is
approximately equal to the comoving peculiar velocity divided by
$H\Omega^{0.5}$.  
This scaling with $\Omega$ is not surprising since
the virial theorem implies that the $rms$ value of the physical
velocities in virialised regions with a given density contrast has
similar scaling with $\Omega$.
Note however that the scaling is only approximate
since the density profile
of virialised objects  has some dependence on $\Omega$.
\section{ $\Omega$ dependence in toy models}

It is clear from the form of (\ref{eulernom}) that the source term
which drives the evolution is larger for lower $\Omega$. Therefore we
expect to see more evolved clustering in a low $\Omega$ universe than
in an $\Omega=1$ universe with the same initial conditions and linear
normalisation.  It is instructive to investigate the effect of the
term $\epsilon$ in cases of special symmetry. Consider first the
spherical expansion or collapse before the occurrence of shell
crossing. This case illustrates the effect of of changing $\Omega$ in
the before shell-crossing.  In cases of special symmetry, we find it easier
to solve directly for $\delta$ and $\theta$ rather than for $\vg $ and
$\vtheta$.  Therefore we take the divergence of (\ref{eulernom}) and
use the Poisson equation to obtain
\begin{equation}
\frac{\dd \theta}{\dd \tau}-\theta-\Pi^2-
\frac{3}{2}\left[1+\epsilon 
(\Omega)\right]\left(\delta -\theta \right)=0 , \label{raych}
\end{equation}
where $\Pi^2=\sum_{i,j}(\partial_{x_i}\beta_j)^2=\theta^2/{\cal N}$
with ${\cal N}=1,2$ and $3$ at the centers of configurations with
planar, cylindrical and spherical symmetry, respectively.  
Therefore, in the spherical top-hat model, the equations (\ref{raych}),
(\ref{contnom}) together with the equations relating $a$, $\tau$,
$\Omega$ and $f$ are sufficient to determine the evolution of the
quantities $\theta$ and $\delta$.  For $\Lambda=0$, the spherical
collapse model can be solved analytically (e.g. Peebles 1980) if the
initial peculiar velocity is neglected. Here we numerically integrate
the equations (14) and (17) under the initial conditions,
$\delta_{i}=\theta_{i}$ with \(|\delta_{i}|\ll 1\), where the
subscript $i$ refers to quantities at the initial time.  These initial
conditions are realistic as they arise naturally in linear theory
(e.g. Peebles 1980).  

\begin{figure}
\centering
\mbox{\psfig{figure=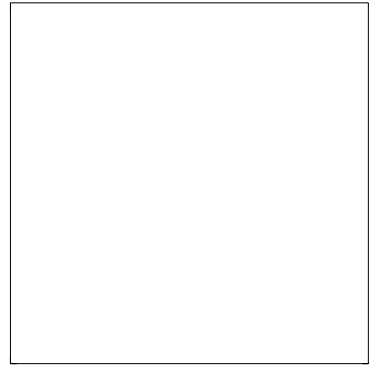,height=6.0cm,bbllx=400pt,bblly=196pt,bburx=10pt,bbury=800pt}}
\caption{The quantities $\delta$ and $\theta$ 
 versus the time $\tau$ for a positive perturbation for various
values of $\Omega_0$ and $\lambda_0$ as indicated in the figure.
The upper steeply rising and the lower curves, respectively, correspond
to bound and unbound perturbations in the $\Omega_0 =0.2$ cases. 
The values of $\delta$ at the turnaround radii of the bound perturbations
are $4.5$, $6.7$ and $11.5$ for ($\Omega_0,\lambda_0$)=($1.,0$), 
(0.2,0.8) and (0.2,0) respectively.
}
\end{figure}

\begin{figure}
\centering
\mbox{\psfig{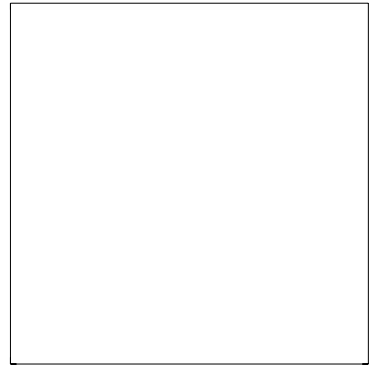}}
\caption{The same as figure 1 but for negative perturbations. The
$\delta$ curves are almost indistinguishable.
}
\end{figure}

\begin{figure}
\centering
\mbox{\psfig{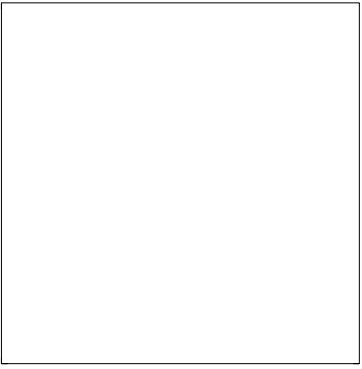}}
\caption{The density parameter versus the time $\tau$ for 
an open universe (dashed line) and
a flat universe with a cosmological constant (solid line).
}
\end{figure}
In figures 1 and 2 we show the density contrast
versus the time $\tau$ for positive ($\delta_i>0$) and negative
($\delta_i<0$) tophat perturbations for three background cosmologies:
($\Omega_0,\lambda_0$)=($1,0$), ($0.2,0.8$) and ($0.2,0$).  Figure 1
shows curves for two values of the initial density contrast
corresponding to bound and unbound perturbations for the
$\Omega_0=0.2$ cosmologies.  Although the equations were integrated
from $\tau=-5.4$ to $0$, for the sake of clarity, figure 1 shows
results only for $\tau>-1$.  For bound perturbations, the growth of
$\delta$ and $\theta$ is fastest in the open universe case,
($\Omega_0=0.2$, $\lambda_0=0$).  However, significant deviations
appear only when $\delta$ is larger than $10$ or so.  This is
consistent with the work of Peacock and Dodds (1996) who found that
nonlinear effects in the evolution of power spectra in N-body
simulations are stronger in an open universe than in a flat
$\Omega+\lambda$ universe of the same $\Omega$.  Although the
$\Omega+\lambda=1$ case shows more rapid evolution than the $\Omega=1$
case, the corresponding curves are very similar even when the
densities are larger than their values at turnaround.  The reason for
this is clear from figure 3 which shows $\Omega$ versus $\tau$. For
$\lambda_0\ne 0$ we see that $\Omega$ is almost unity until relatively
late times. Therefore, until late times, the evolution of $\delta$
and $\theta$ is very similar to the $\Omega_0=1$ case.  For positive
unbound and negative perturbations, the effect of the cosmology on the
evolution of $\delta$ is almost negligible. The $\theta$ curves show
some differences. In voids, $\theta$ grows more slowly in the low
$\Omega_0$ models once the density contrast approaches $-1$.  Since
perturbations in flat universes with a cosmological constant evolve
similarly to those with $\lambda=0$, we do not discuss the case
$\Lambda \ne 0$ further.

The top-hat model can be used to evaluate how 
the variance of an evolved generic density fluctuation field depends on
$\Omega$. We require here that the density field is smoothed on large
enough scales such that shell-crossing is removed.  
Curves of  $\delta(\Omega<1)$ versus 
$\delta_1\equiv \delta(\Omega=1)$ 
satisfy the following (empirical) relation
\begin{equation}
\delta(\Omega)=\delta_1
\exp\left[\frac{\delta_1}{\Delta(\Omega)}\right] 
, \label{domega}
\end{equation} 
where
\begin{equation}
\Delta=\frac{85}{\epsilon(1+\epsilon)} .  \label{fit}
\end{equation}
This relation works remarkably well for $0.1<\Omega<1$ and
$\delta_1<400$.  For generic configurations we assume that the
relation (\ref{fit}) is still valid. However we should take into
account the fact that the ``dimensionality'' of the collapse affects
the amplitude of the $\Omega$ dependence; for example in the one
dimensional collapse, before shell crossing, the equations are free of
$\Omega$. Therefore, we replace $\Delta$, in (\ref{fit}), with
$4\Delta/({\cal N}-1)^2$ where ${\cal N}$ is the ``dimensionality'' of
the collapse at each point in space.  Other than for purely symmetric
configurations, the quantity $\cal N$ is somewhat ambiguous. One
possibility is to define it at any point in space in terms of the
eigenvalues of the initial velocity deformation tensor, say, as the
ratio of the square of the sum of the eigenvalues to the sum of their
squares.  Nevertheless, for our purposes it is not crucial to specify
the form of $\cal N$ and we simply treat it as a factor which depends
on the local topology of the density field.  If $P$ is the probability
distribution function of the field $\delta_1$ evolved with $\Omega=1$,
then it can be shown that the variance, $\sigma^2(\Omega)$, of the
field $\delta(\Omega)$ is given by
\begin{equation}
\sigma^2(\Omega)=\int P(\delta_1)\frac{1+\delta_1}{1+\delta(\Omega)}
\left[1+\delta(\Omega)\right]^2 {\rm d}\delta_1 -1 , \label{prob}
\end{equation}
where we have introduced the ratio $(1+\delta_1)/(1+\delta) $ to
account for the change in the volume element as a result of the change
in density. 
By expanding (\ref{domega}) to 
third order  in $\delta_1$ and  substituting the 
result in (\ref{prob}) we
find
\begin{equation}
\frac{\sigma^2(\Omega)}{\sigma^2_1}=
1+\frac{1}{a_N \Delta}+S_3\left(\frac{1}{2}+
a_N \Delta  \right) 
\frac{ \sigma^2_1}{a_N^2\Delta^2} , \label{sigom}
\end{equation}
where $\sigma_1^2$ is the variance of the field $\delta_1$ and
$a_N=4/({\cal N }-1)^2$ where $0<{\cal N}<3$ is some number describing
the dimensionality of typical collapse configuration. Note that, in
deriving the last relation, we have neglected any local correlation
between $\cal N$ and $\delta_1$.  The factor ${\cal}$ should depend on
the power spectrum of the initial fluctuations and it can be
determined empirically from N-body simulations.

 Consider now the effect of changing $\Omega_0$ on the dynamics in
shell-crossing regions in the case of 1-dimensional collapse.  This case
is particularly instructive since solutions ($\vtheta=\vg$) to the
1-dimensional equations of motion are fully independent of $\Omega$
until the occurrence of shell-crossing. Unfortunately, even in the
simple one dimensional collapse, we have no analytic solutions in
shell-crossing regions.  Therefore we first use the  Zel'dovich
solution until the formation of the first singularity. Then we switch
on to a one-dimensional N-body code to move particles according to
(\ref{eulernom}) in the shell-crossing phase.  The initial density field
we choose is $\delta_i\propto \cos(x)$.  Results of simulations with
$\Omega_0=1$ and $\Omega_0=0.2$ without a cosmological constant are
shown in figure 4.  The density profile (upper panel) is more
concentrated in the open than in the flat case. This is similar to the
behavior of density perturbations in the  tophat model
discussed above.  The distribution and velocities of particles in the
open model seem to be more evolved in time than in the flat model.

\begin{figure}
\centering
\mbox{\psfig{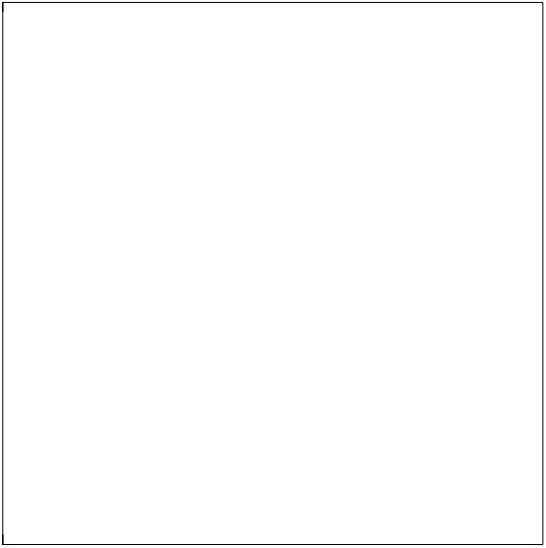}}
\caption{The density profile (upper panel) and the ``velocities'' 
 (lower panel) in the shell-crossing region in one dimension for $\Omega_0=1$
(solid) and $\Omega_0=0.2$ (dotted) with no cosmological constant. 
The initial density perturbation is a cosine wave symmetric about $x=0$.
}
\end{figure}

\section {$\Omega$ dependence in N-body simulations}

We use N-body simulations  to study 
the $\Omega$ dependence under general
initial conditions. These
simulations are especially useful in orbit 
mixing regions where, according to 
(\ref{eulernom}), the effect is most important.

We ran two  simulations having $\Omega_0=1$ and $\Omega_0=0.29$
respectively. Both simulations were started from the
same initial conditions.  
The initial conditions were generated from  the power spectrum for 
standard CDM with $H_0=50{\rm km ~s^{-1} Mpc^{-1}}$. Each simulation
contained $128^3$ particles in a cubic box of length $60 {\rm h^{-1}Mpc}$.
The simulations were evolved until the linear $rms$ density fluctuations
in a sphere of $800{\rm km ~s^{-1}}$ was 0.5.
Both models have roughly the right small scale power
as measured by the galaxy pairwise velocity, but produce fewer 
 rich clusters than observed.
However, given the 
scale of our simulations, our choice of the power spectrum and the
normalisation is  appropriate for our purposes.
A model with a higher normalisation would result in too much merging of
smaller objects into a few larger objects. Choosing a steeper
power spectrum leads to a similar effect.

The simulations were run using  a modified  parallel version 
(MacFarland \etal 1997) of Couchman's P$^3$M code 
(Couchman \etal 1995) which uses
explicit message passing. The simulations had a softening parameter
of $13.2\%$ the mean particle separation and a mesh size
of 512 in one dimension.
They were run using 
64 processors on 
the CRAY T3E supercomputer at the
Computer Center of the Max Planck Society (RZG), Garching.


\begin{figure}
\centering
\mbox{\psfig{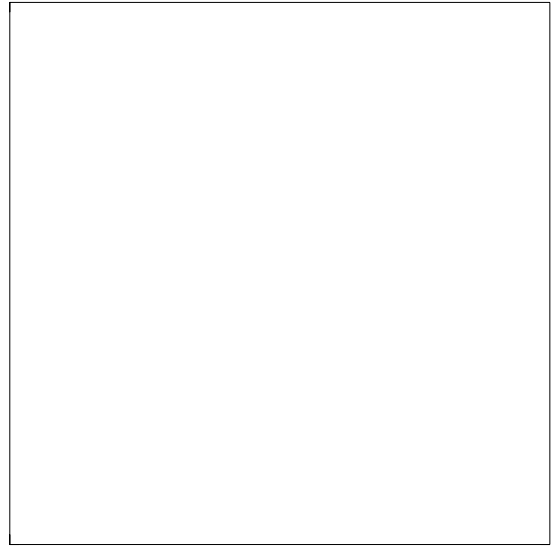}}
\caption{The particle distribution in the low $\Omega$ (right)
and $\Omega=1$ (left) simulations.
Slice thickness is $1{\rm h^{-1}Mpc}$. The lower panels focus on the
group of ``clusters'' appearing near the centers of the upper panels.
}
\end{figure}

\begin{figure}
\centering
\mbox{\psfig{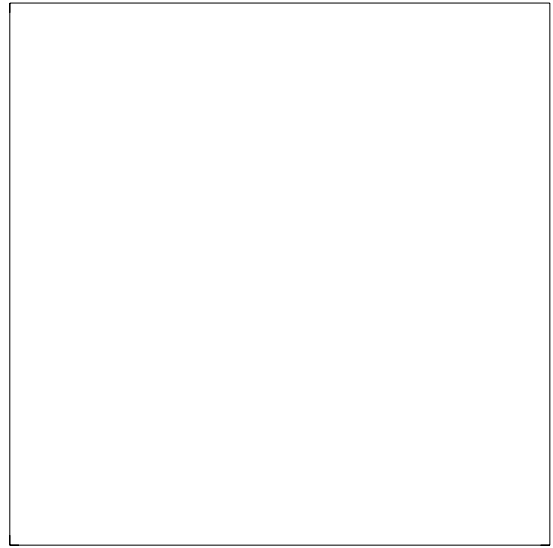}}
\caption{The two point correlation functions for the particle distribution
in the two simulations. 
}
\end{figure}

Figure 5 shows
the particle distribution in a slice of thickness $1{\rm h^{-1}Mpc}$,
in the two simulations. The left and right panels correspond to the
flat and open model respectiveley.  The lower panels zoom in on the
``clusters'' seen near the centers of the upper panels. On large
scales (upper panels) the two simulations are remarkably similar. Some
differences can be spotted in the lower panels.  Clusters in the low
$\Omega$ simulation appear to be more concentrated and evolved.  The
differences between the the two simulations seem to be negligible on
scales larger than $1 {\rm h^{-1}Mpc}$. Indeed the $rms$ value of the
difference between the positions of the same particles in the two
simulations is $0.25{\rm h^{-1}Mpc}$ and the largest difference is
less than $1.5{\rm h^{-1}Mpc}$.  The correlation functions for the two
simulations, plotted in figure 6, confirm the visual impression from
figure 5. The correlation functions differ only on scales smaller than
$1.6{\rm h^{-1}Mpc}$. On ``cluster'' scales $\lsim 0.6 {\rm
h^{-1}Mpc}$ the low $\Omega$ correlation function is larger and, on
scales $0.6-1.6{\rm h^{-1}Mpc}$ roughly corresponding to the scale of
infall regions around clusters in the simulations, it is smaller than
the  correlation function in the flat model.
\begin{figure}
\centering
\mbox{\psfig{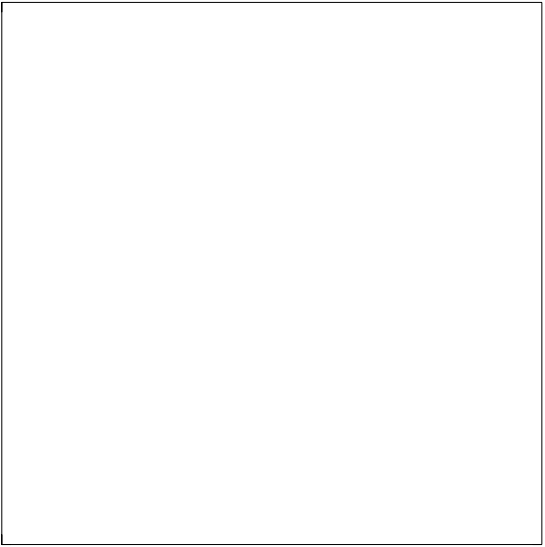}}
\caption{Densities in the open model vs. densities in the 
the flat model. The lower left panel shows densities after the CIC
interpolation on a cubic grid of mean particle separation cell size.
The other panels show CIC densities smoothed with a top-hat 
window of width $R_s$ as indicated in the plot.
}
\end{figure}

\begin{figure}
\centering
\mbox{\psfig{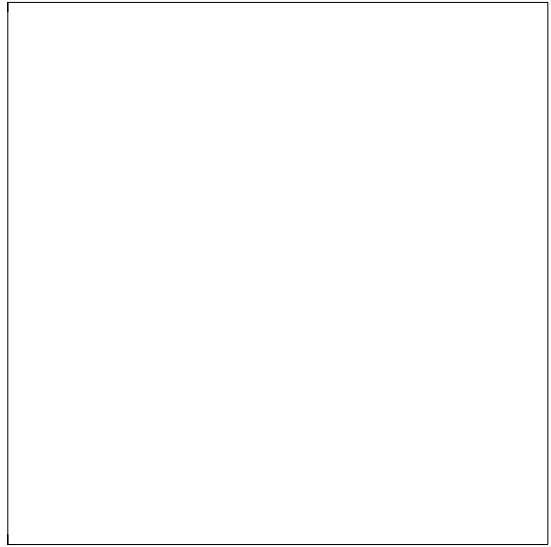}}
\caption{The same as figure 6 but for peculiar velocities.
The velocities in the open model were scaled by the factor
 $f(\Omega_0)=0.477$.
}
\end{figure}
We now quantify the differences between the density and velocity
fields.  We first use the cloud-in-cell (CIC) interpolation scheme to
evaluate the density and velocity fields on a cubic grid of cell
length equal to the mean particle separation. This produces a mass
weighted average velocity on the grid points. We then further smooth
the resultant density and velocity maps with a top-hat filter.  In
figure 7 we plot the densities in the open vs those in the flat model
for 4096 randomly chosen grid points.  Even with only CIC smoothing on
the grid scale, the correlation between the two density fields is very
tight. 
 For densities larger than $10$ or so the densities in the open
model are larger.  The scatter almost vanishes when the density fields
are smoothed with a top-hat window of width $>200{\rm km/s}$.  Note
however that for moderate densities ($0<\delta<5$), the densities in
the flat model are slightly larger than the open model.  This is not
surprising since the general tendency is that matter flows out of
regions with moderate densities into higher density regions.  And
since the open model is slightly more evolved, these moderate density
regions are somewhat less dense in the open than in the flat model.
In table 1 we list the values of the $rms$, $\sigma_{\delta}$, and the
reduced skewness, $S\equiv <\delta^3>/\sigma^4_{\delta} $ of the
density fields as a function of the smoothing scale.
\begin{table}
\caption{Moments of the density field in the two simulations after CIC
and top-hat smoothing of width,$R_s$, expressed in km/s.}
\begin{tabular}{|c|c|ccc} \hline
& \multicolumn{2}{c}{$\Omega_0=1$}
&\multicolumn{2}{|c|}{$\Omega_0=0.29$} \\ \hline
&\multicolumn{1}{c}{$\sigma_{\delta}$}&\multicolumn{1}{c}{$S$}&\multicolumn{1}{c}
{$\sigma_{\delta}$}&\multicolumn{1}{c}{$S$}
\\ \hline
CIC &    7.331 &6.222 &8.816&6.374 \\ \hline
$R_s=50$ & 6.313 & 5.934 &7.397 & 6.154\\ \hline
$R_s=100$ & 3.331 & 4.855 & 3.603 & 5.035 \\ \hline
$R_s=200$ & 1.693 & 3.720& 1.752 & 3.833 \\ \hline
$R_s=400$ & 0.878 &3.060&  0.891& 3.116 \\ \hline
$R_s=800$ &0.440  &2.549& 0.443 & 2.584 \\ \hline
\end{tabular}
\end{table}
It appears, from the table, that the dependence of $S$ on $\Omega$ is
stronger than what is predicted from second order perturbation theory
(Bouchet \etal 1992).  That theory is, however, valid only for
$\sigma<1$.  We can use table 1 to determine ${\cal N}_{eff}$ in
(\ref{sigom}) which relates $\sigma_\delta$ in an open universe to
that in an $\Omega=1$ universe. A comparison of (\ref{sigom}) with
table 1 suggests that ${\cal N}_{eff}\approx 2$. This value is
reasonable since non-linear collapse configurations are likely to have
pancake-like shapes. Recall that the simulations were stopped when the
the linear value of of $\sigma_\delta$ smoothed with a top-hat filter
of width $R_s=800{\rm km/s}$ was 0.5.  The actual value
computed from the simulation is very close to $0.44$ in the two
simulations.  Thus, even though nonlinear effects are clearly
important, the difference between the $\sigma_8$ in the open and flat
simulations is negligible. 
We now consider the evolved velocity fields. Figure 8 compares one of
the components of the velocity fields in the two simulations.  The
velocity fields in the open model are scaled by the factor
$f(\Omega_0)$. Even for large velocities and small smoothing widths,
the velocity fields in the two simulations seem to be related by the factor
$f$. A close inspection of the scatter plot for $R_s=400{\rm km/s}$
reveals that the slope of the regression of $v/f$ on $v_1$ is slightly
less than unity.  This is because
large velocities are generally
associated with strong nonlinear effects which tend to spoil the
scaling by $f$.

We mentioned at the end of section 2, that the motion of particles in
bound objects is independent of $\Omega$ in terms of a time variable
which corresponds to a velocity which is the peculiar comoving
velocity divided by $\Omega^{0.5}$.  Therefore we expect the velocity
dispersion ($rms$ velocity) in groups of particles identified in the
simulation to scale, approximately, like $\Omega^{0.5}$. To test this
conjecture, we have used a friends-of-friends (f.o.f) algorithm
(kindly supplied by A. Diaferio) to identify groups in the
simulations.  We then computed the one-dimensional velocity dispersion
of particles in each group.  Figure 9 (upper panel) shows the mean
velocity dispersion in groups as a function of the number of particles
they contain. Velocities in the plot are scaled by $\Omega_0^{0.5}$
for the open model.  It seems that this scaling works well.  Given the
uncertainties in identifying group members by the f.o.f algorithm, the
deviations from this scaling for large groups are not significant.  We
conclude that while the (smoothed) average velocity of particles
scales like $f(\Omega)$, the $rms$ velocity, roughly, scales like
$\Omega^{0.5}$.  It is interesting to compare the abundance of groups
in the two simulations.  The lower panel of figure 9 is a plot of the
abundance as a function of the number of particles for the two
simulations.  The abundance of groups is slightly higher in the open
model. This is easy to understand, because groups in the open model
are tighter than groups in the flat model, the f.o.f algorithm
naturally assigns more particles to them.  Note that observations
naturally provide the abundance of groups as a function of the
mass. Since the mass of groups with the same number of particles is
proportional to $\Omega_0$, abundance curves, when plotted versus the
mass, look significantly different in the two simulations.

\begin{figure}
\centering
\mbox{\psfig{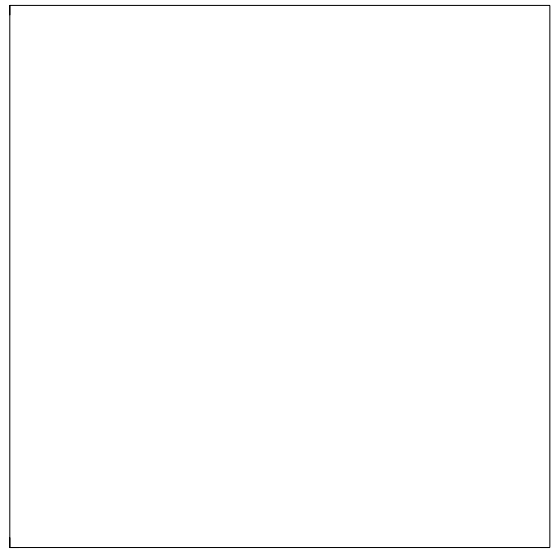}}
\caption{{\it Top}: The mean 1-D velocity dispersion ($rms$ velocity
deviations) divided by $\Omega_0^{0.5}$ in groups vs. the number of
particles they contain for the two simulations.  The lengths of the
errorbars give the 1-sigma scatter about the mean for the flat model.
The scatter in the open model is similar.  {\it Bottom}: The abundance
of groups vs. the number of particles they contain. Groups are
identified using a f.o.f algorithm.}
\end{figure}

\section {Summary}

We have shown that gravitational dynamics of a pressureless fluid in
an expanding universe is almost independent of the cosmological
parameters. According to the equations of motion, expressed in terms
of the linear growing mode, the final structure in a low $\Omega$
model, with or without a cosmological constant, is more evolved than in
a flat universe. We used toy-models and N-body simulations to
investigate the effect of changing the cosmological background on the
evolution of fluctuations and, in particular, on the final velocity
and density fields.  The present density, when smoothed on scales
slightly larger than cluster scale, is almost insensitive to the
cosmological background. The background can affect the structure of
bound objects (or halos). Halos, characterised by the same ratio of
mass to background density, are more centrally concentrated in an open
than in a flat universe. However, this effect is weak and is likely to
depend on the initial power spectrum.  On the other hand, the
amplitude of the comoving peculiar velocities of particles strongly
depends on $\Omega$. It is remarkable that the smoothed nonlinear
velocity field scales with the growth factor, $f$, just as it does in
linear theory.

Since $f$ depends very weakly on the cosmological constant, the final
velocity field is mainly sensitive to $\Omega$.  Therefore the
observed peculiar velocity and density fields in the nearby universe
contain information only on $\Omega$ (Lahav \etal 1991). However,
constraints on both $\Omega$ and $\Lambda$ can, in principle, be
obtained by measuring the clustering amplitude at different redshifts,
for example via the correlation function (Lahav \etal 1991). As we
have shown, it is rather difficult to detect signatures of the
cosmological background in the structure of density
fields. Fortunately, observations provide the distribution of galaxies
in redshift space.  Thanks to the strong dependence of the velocity
field on $\Omega$, the anisotropy of clustering in redshift space can
provide a measure of $\Omega$.  However, such estimates of $\Omega$
involve an assumption on the relationship between the distribution of
galaxies to that of the dark matter. Estimates of $\Omega$ independently
of the galaxy distribution can be obtained from the observed peculiar
velocity field alone (c.f. Dekel 1994). This makes peculiar velocity
catalogs a very powerful too to constrain the cosmological model. It
is especially important for future peculiar velocity measurements to
aim at larger sky-coverage and denser sampling rate.

\section*{Acknowledgments}

We especially would like to thank Ravi Sheth for many useful comments,
and Tom MacFarland for his
valuable contributions to running and improving the parallel P3MN-body code.
We also thank Simon White for useful discussions, and  Antonaldo Diaferio for allowing the use of his group finding
code.

\protect

\end{document}